# Can Voice Assistants Sound Cute? Towards a Model of Kawaii Vocalics

**KATIE SEABORN**

*Tokyo Institute of Technology*
*Tokyo, Japan*

**SOMANG NAM**

*Toronto Metropolitan University*
*Toronto, Canada*

**JULIA KECKEIS**

*TU Berlin*
*Berlin, Germany*

**TATSUYA ITAGAKI**

*Tokyo Institute of Technology*
*Tokyo, Japan*

**ABSTRACT:** The Japanese notion of "kawaii" or expressions of cuteness, vulnerability, and/or charm is a global cultural export. Work has explored kawaii-ness as a design feature and factor of user experience in the visual appearance, nonverbal behaviour, and sound of robots and virtual characters. In this initial work, we consider whether voices can be kawaii by exploring the vocal qualities of voice assistant speech, i.e., kawaii vocalics. Drawing from an age-inclusive model of kawaii, we ran a user perceptions study on the kawaii-ness of younger- and older-sounding Japanese computer voices. We found that kawaii-ness intersected with perceptions of gender and age, i.e., gender ambiguous and girlish, as well as VA features, i.e., fluency and artificiality. We propose an initial model of kawaii vocalics to be validated through the identification and study of vocal qualities, cognitive appraisals, behavioural responses, and affective reports.

## 1   Introduction

When we think of Japanese culture, we often think of cute and charming characters—mascots and characters from anime, manga, video games, and other mass media. The impressions of cuteness, charm, sweetness, and adorableness that these characters invoke is called "kawaii" in Japanese [9,16,18,25,29,30,34]. The aesthetic of kawaii can be found across an array of media, from Pikachu and other "pocket monsters" in the Pokémon video game and anime series to the titular transmedia character Hello Kitty and friends licensed by Sanrio, found embedded within products ranging from toys to stationary to household items to theme parks. Kawaii as an attribute of agents (human or not) and objects is typically linked to visual appearance, static or animated [9,16,30,34], but may be portrayed through other modalities, as well.

Indeed, researchers working in human-computer interaction (HCI), human-robot interaction (HRI), and human-agent interaction (HAI) have started exploring the "kawaii-ness" of interactive agents that feature a range of modalities and communicative capacities. So far, robots, virtual characters, and artificial agents have taken centre stage, with most focusing on visual appearance [4–6,9,30] but emerging work on movement and nonverbal behaviours [48], nonverbal sounds and melodies [9,51], touch [35], conduct [24,27], and voice [24]. This shift towards exploring nonvisual modes of expressing kawaii in interactive agents echoes an increase in HCI work on voice-based agents without a visible or perceivable body [43]. Voice assistants (VAs), in particular, have experienced a surge in popularity commercially as well as within HCI and adjacent spaces [1,10,12,17,43]. Since most VAs take the form of a smart speaker, i.e., LINE's Clova or Amazon's Alexa, a speaker embedded within a space, i.e., smart vehicles, and/or an abstract visualization, i.e., Apple's Siri, sound may be the best or only medium through which kawaii can be expressed. Yet, this is virtually unexplored.

The phenomenon of "cuteness" and kawaii was first approached as an object of study [19,31,34] by drawing on Lorenz's notion of "Kindchenschema" or "baby schema" [23]. This refers to the visual properties of young animals that are theorized to stimulate a care response: a large head and eyes compared to the body, roundness and softness in shape and texture. Kawaii is also said to carry gendered associations that intersect with age: that it is a girl's thing [16] and/or a socially constructed ideal of femininity as vulnerable, unthreatening, compliant, and immature [47]. Still, Nittono and colleagues [29–33,35] uncovered limitations in operationalizing kawaii as baby schema and gendered within the Japanese sociocultural context: objects without anthropomorphic characteristics, such as flowers, dessert, and accessories, were also perceived as kawaii. These findings were organized into a two-layer model of kawaii: social value and emotion [29]. Kawaii was also distinguished from "cuteness" based on agedness, with smiling older adults deemed kawaii [29,31,32]. This is not unprecedented; as Shiokawa wrote over two decades ago, kawaii can describe "likable personal quirks in not-so-young folks, especially the elderly" (p. 93). Qualitative work recently captured this under the term "otona-kawaii" or "adult-cute" to explain the social acceptance of cute behaviour in older people [22]. This raises the question of whether the agelessness and genderlessness of kawaii applies to voice or is only a matter of body and behaviour.

In this preliminary work, we used the two-layer model of kawaii [29] to explore kawaii in the voice of bodyless interactive agents, i.e., VAs. We asked: *RQ1. Can computer voice invoke a sense of kawaii in the absence of visual indicators and the presence of anthropomorphic cues?* And *RQ2. Does computer voice agedness and genderedness affect perceptions of kawaii?* To this end, we carried out an initial user perceptions study of kawaii in the vocal qualities of a range of simulated VAs. We contribute initial findings on kawaii voice, diverging from the "agelessness" findings for visual stimuli [29,31,32] and providing nuance on perceptions of gender as an intersection with age in kawaii voices, expanding the purview of the social layer to include social identity. We also identify the VA features of fluency and artificiality as influential. We provide a preliminary model for launching a new area of study within the VA domain: *kawaii vocalics*. We offer this work as a first step towards recognizing and understanding what voice qualities invoke a sense of kawaii.

## 2   Related Work

### 2.1   KAWAII VOCALICS? VOCAL QUALITIES OF COMPUTER VOICE

Kawaii is typically operationalized as a *visual* and/or *physical* property, i.e., a matter of appearance that can be visually perceived. Yet, there is no reason to limit kawaii to being a visual phenomenon, especially given how it intersects with and may rely on other modalities, such as sound. Indeed, we raise this possibility as a key driver of this work and seek to extend the notion of kawaii to the auditory modality as a sound- (or voice-)based stimulus.

*Vocalics*, or the *paralinguistic* qualities of voice, refer to "meta" vocal qualities beyond words, such pitch, volume, rate of speech, verbal fillers, and timbre, that convey information beyond speech content, notably emotion, personality, and social qualities, such as gender and age [38]. A variety of vocal qualities have been explored in computer voice over the past several decades [10,43,46]. Yet, kawaii voice is virtually unexplored. Cheok [9] found that high-pitched melodies were rated as most kawaii. Lv et al. [24] explored "cute" voices for VAs by demarcating cuteness as tone of voice, i.e., vocalics, and language style, i.e., speech patterns, finding that a combination of the two was rated most cute. Zhang et al. [51] found a preference for quiet and cute-sounding, high-pitched consequential sounds for an industrial robot arm. Kumagai [20] found that sounds produced when we touch our lips together, i.e., bilabial sounds, were associated with kawaii. The dearth of work on kawaii vocalics may be surprising, especially considering Japan's culture of singing voice synthesizers, notably Hatsune Miku (2000), a Vocaloid software "voicebank" created by Crypton Future Media. Yet, the consensus is that humanlikeness is a key ingredient [3,10,43]. This led to our first hypothesis, that:

> H1. Perceptions of kawaii voice will depend on perceptions of low artificiality and high anthropomorphism.

## 2.2   OPERATIONALIZING KAWAII AS A SOCIOCULTURAL PHENOMENON

Kawaii is not a new concept within the cultural landscape of Japan. Its origins are somewhat unclear, with roots in the 11th century novels of "Genji monogatari" ("The Tale of Genji") by Lady Murasaki [47] and "Makura no sōshi" ("Pillow Book") by Sei Shonagon [16]. Etymologically, it emerged through a combination of different linguistic sources, from a phrase for blushing to the morphemes for "dazzling" as well as "embarrassment" or "awkwardness" [47]. Lady Murasaki, for instance, used the word to refer to the pitiable, helpless nature of certain characters to invoke a sense of sympathy in the reader, a meaning that was eventually adopted in the term kawaisō [47]. Kawaii was carried forward into the Edo period (1600s-1800s) by Kabuki performers and novelists and reinvigorated in the last century by cultural trends, like girl's writing [9], and mass media, especially manga (comic) artists and animators, notably Harada Osamu, Yuko Shimizu, and Rune Naito, and companies like Sanrio [9,16]. Still, its flavour has changed over time. Shiokawa deconstructs kawaii as a mere descriptor of cuteness or piteousness, arguing that it is an ever-shifting, liminal construct, purposefully vague and ill-defined so as to fit individual circumstances and modes of expression [47]. This is part of the appeal of kawaii: not being tied to uncontrollable factors, like the face we are born with, allows it to be adopted by anyone. Indeed, kawaii may best be understood in terms of what qualities it does *not* invoke, i.e., lack of a threat response. Inuhiko notes that the juxtaposition of "piteousness" against impressions of cuteness and charm also invokes a sentiment of the grotesque [16]. We may at first be surprised by Sailor Moon's gigantic eyes, but over time we come to view these features positively.

Nittono and colleagues [28,29,31–33] recognized that, as a sociocultural phenomenon, kawaii could also be viewed as a psychological phenomenon. After a series of foundational studies involving Japanese and non-Japanese people [33] and people of all ages and genders [28,32], Nittono discovered that this liminal, shifting, and somewhat contradictory notion of kawaii set it apart from other cultural notions of cuteness and called into question dominant understandings of its nature to Japanese people [29]. One layer is *emotion*, specifically positive affective responses elicited by kawaii stimuli. The other layer is *social value*, referring to kawaii as a medium of sociality. For example, "kawaii spirals" are where a kawaii stimulus leads to reciprocal smiles and positive emotions among two or more people [29].

Notably, the role of social identity factors, like age and gender, are absent from this model. Yet, how kawaii has been historically conceptualized and studied suggests that we should. We seek to extend the social layer to accommodate perceptions of two social identity factors highlighted so far. The first is age. Nittono et al. [32] found that images of older adults smiling also invoked a sense of kawaii. Lieber-Milo [22] explored the notion of "otona-kawaii" (adult-kawaii) and found that Japanese adults had heard of it or were open to the concept. This leads us to our first hypothesis:

> H2. *Perceptions of voice age will not be linked to kawaii, i.e., kawaii voice is an ageless phenomenon.*

As a psychological and sociocultural construct, kawaii has been located at the intersection of age and gender, notably as a feature of babies and children, but also young women [16,47]. For this reason, some have criticized the concept as a Confucian mode of infantilizing women and femininity [47]. Others have been said to dismiss the concept as a "girl's thing" [16,47], which itself indicates a level of social bias. Shiokawa provocatively argues that kawaii has been molded and reconstituted over time, especially as women gained more power and influence in mass media [47]. Still others have approached kawaii as gender-neutral [29,49]. Still, kawaii has been painted feminine in broad strokes. Thus, we consider:

> H3. *Perceptions of voice gender will be linked to kawaii in terms of femininity, i.e., kawaii is gendered feminine.*

We must also consider the intersections when it comes to matters of social identity [11,26,37,40]. While work on social identities in HCI and adjacent spaces is growing—in populations [8,21,40] and in design materials, such as personas [26], creations [42], and underlying technologies [7]—we must go beyond analyzing multiple factors in isolation by considering how these factors intersect in ways that may be difficult or impossible to tease out. When it comes to kawaii, we may specifically expect a link between perceptions of *younger* age categories and *femininity.* We thus hypothesize:

> *H4. Perceptions of voice age and gender will intersect such that younger-sounding feminine voice perceptions will be linked to user perceptions of kawaii, e.g., stereotype of kawaii as a girl's phenomenon.*

## 3  Methods

We conducted an online user perceptions study in line with the research design of visual, e.g., [31,32], and voice-based, e.g., [2,3], user perception studies. This study was folded into a larger study about user perceptions of computer voice age. Our protocol was registered before data collection on December 22$^{nd}$, 2022 via OSF[1].

### 3.1  PARTICIPANTS

Participants (n=94, women n=53, men n=37, another gender or N/A n=4) were recruited through Yahoo! Crowdsourcing Japan on December 23$^{rd}$, 2022. Most were aged 45-54 (n=34) or 35-44 (n=29), with some younger (18-34 n=15) and older (55-74 n=13). While most were non-users of VAs (n=58), many were daily or weekly users (n=25). Four used VAs once a month and four used to but did not anymore. Six responses were removed because they were incomplete. Participants were paid in accordance with the participant pool at roughly 1200 yen per hour, equating to about 400 yen for 20 minutes.

### 3.2  PROCEDURE

Participants listened to short (10-15-sec.) clips of computer voices simulating utterances by VAs; refer to 3.3. They rated these voice stimuli based on vocal and social qualities; refer to 3.4. They were presented in a random order to counter novelty and order effects [41]. Participants provided demographics on the last page. The study took about 20 min.

---

[1] https://osf.io/eb7zx

## 3.3 MATERIALS

### 3.3.1 Voice Stimuli.

We used eleven voices from CoeFont[2], a Japanese TTS provider: 林アナ, よろこびおじさん, 岡田斗司夫, 小夜 SAYO[β], なな 9歳, 怒るおじさん, 高橋 俊輔, あさのゆき, 淑江おばあちゃん, けんしん, and さくら. We also used three novel Japanese older adult TTSs: an older woman, an older man with a lower-pitched voice, and an older man with a higher-pitched voice; clips are in Supplementary Materials. Pilot tests indicated that the voices carried attributes related to a range of ages and genders, potentially varying perceptions of kawaii-ness by these factors.

### 3.3.2 Speech Content.

We drew inspiration from the scripts by Baird et al. [3], translating three phrases into Japanese: "Thank you" as 「ありがとうございます」, "How are you?" as 「おげんきですか」, and "I love you" as 「あなたを愛しています」.

## 3.4 MEASURES AND INSTRUMENTS

All instruments used a 5-point Likert scale. Item order was randomized for each voice to avoid order effects [41]. All items were translated into Japanese by a native speaker and back-translated with an advanced speaker native in English.

### 3.4.1 Kawaii Perceptions.

We included one item simply asking for a rating of kawaii-ness. We operationalize kawaii-ness as a mean greater than 3.5 (skewed towards agreement on kawaii-ness) and a median of 4 or above (nominal agreement).

### 3.4.2 Perceptions of Anthropomorphism/Humanlikeness, Artificiality, and Fluency.

We used the 1-item humanlikeness scale from Baird et al. [3], dividing the poles (humanlike and artificial) into separate items to match the structure of the other items in the survey. We also asked about language fluency as a potential confounding variable for anthropomorphism, given the differing technical quality of the TTSs we used [43,50].

---

[2] https://CoeFont.cloud

*3.4.3 Age Perceptions.*

Agedness was captured in a nominal scale comprised of infant/baby (0-2 years), child (3-12 years), teenaged (13-19 years), adult (20-39 years), middle-aged (40-64 years), older adult (65+ years), and ageless.

*3.4.4 Gender Perceptions.*

Genderedness captured in a nominal scale comprised of feminine, masculine, aspects of both, and neither, with the last two operationalized as gender ambiguous. Participants were also free to enter another option or description.

## 3.5 DATA ANALYSIS

We created descriptive statistics for each voice as well as voices grouped by perceived age (younger, middle, older) and gender (masculine, feminine, ambiguous). Shapiro-Wilks tests showed non-normal distributions, so we used non-parametric statistics, e.g., Chi-squares, Kruskal-Wallis H tests, and Kendall's tau-b correlations. Qualitative data, e.g., gender descriptions, were summarized using a conventional content analysis approach [14] by the third author and checked by the first author; we had no disagreements.

## 4 Results

Descriptive statistics showed that several voices were perceived as kawaii: the "teenaged girl"小夜_SAYO[β] (M=3.6, SD=1, MD=4, IQR=1); the "young girl"なな (M=3.9, SD=0.9, MD=4, IQR=0); the "young boy"けんしん (M=3.7, SD=0.9, MD=4, IQR=1); and the "young girl'さくら (M=3.7, SD=0.8, MD=4, IQR=1). We now consider the hypotheses and qualitative findings.

## 4.1 H1. PERCEPTIONS OF KAWAII VOICE WILL DEPEND ON LOW ARTIFICIALITY AND HIGH ANTHROPOMORPHISM.

A strong, negative correlation was found between Artificial and Humanlike ratings ($\tau b$ = -.640, $p$ < .05). A weak, positive correlation was found between Kawaii and Humanlike ratings ($\tau b$ = .052, $p$ < .05). A positive, moderate correlation was found between Humanlike and Fluent ratings ($\tau b$ = .456, $p$ <.05) and a negative, moderate correlation was found between Artificial and Fluent ratings ($\tau b$ = -.354, $p$ < .05). No others were found. These statistically significant correlations and their directions support the hypothesis: *kawaii voices are the most fluent, humanlike and least artificial.*

## 4.2  H2. PERCEPTIONS OF VOICE AGE WILL NOT BE LINKED TO KAWAII.

A moderate, negative correlation was found between Perceived Age and Kawaii ratings ($\tau b$ = -.547, $p$ < .01). A Chi-Square test found a significant relationship between Perceived Age and Kawaii ratings, $\chi2(16, 1308)$ = 720.91, $p$ < .05, $\varphi$ = .372. A Kruskal-Wallis test indicated a significant difference by age category, $\chi2(5)$ = 635.16, $p$ < .05, with a Dunn's test (Bonferroni corrected) revealing significant differences across all categories except for two combinations: middle aged-older adult, and child-teen: child (M=3.8, SD=.9, MD=4, IQR=1), teen (M=3.6, SD=1, MD=4, IQR=1), adult (M=2.5, SD=1, MD=2, IQR=1), middle-aged (M=1.6, SD=.7, MD=1, IQR=1), and older adult (M=1.7, SD=.8, MD=1, IQR=1). Thus, we cannot accept the hypothesis; *kawaii appears to be an age-based phenomenon*, linked to voice age and favouring youth.

## 4.3  H3. PERCEPTIONS OF VOICE GENDER WILL BE LINKED TO KAWAII IN TERMS OF FEMININITY.

A Chi-Square test found a statistically significant association between Perceived Gender and Kawaii ratings, $\chi2(8, 1307)$ = 337.19, $p$ < .05, $\varphi$ = .359. Follow-up Chi-Square tests (Bonferroni corrected) showed relationships between Perceived Gender and Kawaii ratings for all categories: both (M=3.4, SD=1.2, MD=4, IQR=1.8), feminine (M=3, SD=1.2, MD=3, IQR=2), and masculine (M=1.9, SD=1, MD=2, IQR=1), $p$ < .05. No voice was classified as gender ambiguous across participants. A Mann-Whitney U test indicated that feminine voices (MD=3, IQR=2) were more kawaii than masculine ones (MD=2, IQR=1), $U$ = 73926, $p$ < .05. We can partially accept the hypothesis: people who rated voices as feminine *and* people who rated voices as gender ambiguous tended to provide higher ratings of kawaii.

## 4.4  H4. PERCEPTIONS OF VOICE AGE AND GENDER WILL INTERSECT: YOUNGER-SOUNDING FEMININE VOICES AND KAWAII.

Chi-Square tests found a significant relationship between Perceived Age and Kawaii rating for feminine voices, $\chi2(16, 561)$ = 273.01, $p$ < .05, masculine voices, $\chi2(16, 607)$ = 423.11, $p$ < .05, and ambiguous voices, $\chi2(16, 134)$ = 81.05, $p$ < .05. A Kruskal-Wallis test revealed the same for Perceived Gender, $\chi2(2)$ = 308.078, $p$ < 0.05, with a Dunn's test (Bonferroni corrected) showing significant differences for all Perceived

Genders. Similarly, a Kruskal-Wallis test for voice and gender classifications revealed significant differences in Kawaii ratings, $\chi2$(6) = 723.255, $p$ < .05. A Dunn's test using a Bonferroni correction indicated this for all pairs except for masculine adult-feminine older adult and feminine child-feminine teen: feminine child (M=3.8, SD=.9, MD=4, IQR=1), feminine teen (M=3.6, SD=1.0, MD=4, IQR=1), feminine adult (M=3.0, SD=.9, MD=3, IQR=1), masculine adult (M=2.0, SD=.8, MD=2, IQR=1), feminine older adult (M=1.9, SD=.8, MD=2, IQR=1.3), masculine middle-aged (M=1.6, SD=.7, MD=1, IQR=1), masculine older adult (M=1.5, SD=.7, MD=1, IQR=1). We can partially accept the hypothesis, i.e., *the stereotype of a girl's phenomenon*, but *also for gender ambiguity*.

### 4.5  QUALITATIVE INSIGHTS

Out of 94 participants, 29 (30.9%) provided qualitative comments on at least one voice. Only the "teenaged girl" voice 小夜_SAYO[β] (n=2), the "woman" voice 林アナ (n=1), and the "young girl" voiceなな (n=1) were characterized as kawaii; notably, the "woman" voice 林アナ was not otherwise rated as kawaii (M=3.1, SD=0.9, MD=3, IQR=1). All voices were assigned a range of descriptors related to anthropomorphism, including these three voices. For example, the "teenaged girl" voice 小夜_SAYO[β] was described as "very cute" but also "somewhat fake" and "like a synthesized voice." This suggests that perceptions of voice artificiality mediated, especially dampened perceptions of kawaii.

## 5  Discussion

### 5.1  USER PERCEPTIONS OF KAWAII VOCALICS: GIRLISH, GENDER-AMBIGUOUS, AND ANTHROPOMORPHIC

Kawaii voice perceptions appear to relate not just to social behaviour [29] or pitch and/or speech [9,20,24,51], but also anthropomorphic stimuli "hearable" in the voice, specifically the expected [9,47] social categories of age and gender. Even so, we found an unexpected result: the most kawaii-sounding voices were those perceived by individuals as *gender ambiguous*, having a mix of feminine and masculine qualities. This may be explained by Shiokawa's [47] conception of kawaii as *ambiguous* and interpreted by individuals. Machines can also have qualities of gender neutrality [44] and complex gender ascriptions [45]. In contrast to previous work [32], kawaii voice *does* appear to be linked to perceptions of youth, with no suggestion of agelessness or otona-kawaii [22]. Cognitive appraisals from the qualitative findings confirm this for our corpus. We thus offer a preliminary model of kawaii vocalics (Figure 1). We extend the two-layer model [29] to include the social identity characteristics of age and gender and the VA characteristics of fluency and anthropomorphism.

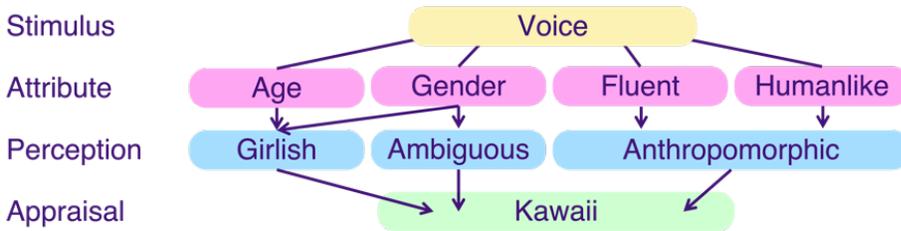

*Figure 1: Preliminary model of kawaii vocalics based on of the two-layer model with social identity and VA extensions.*

## 5.2 IMPLICATIONS: A RESEARCH AGENDA FOR KAWAII VOCALICS

Our preliminary results provide a solid foundation for an emerging model of kawaii vocalics, one that complements and extends its visual origins. We outline the following trajectories for launching its study within voice UX research.

### 5.2.1 Translating Visual Kawaii Attributes to Vocal Attributes.

Translating the visual properties of kawaii to kawaii vocalics may be a simple one-to-one translation of visual to vocal attributes. In effect, this is as a matter of sound symbolism [13] in the context of voice, e.g., are auditory stimuli "round" or "fluffy"? Are attributes such as "bright" and "lovely" perceived through sound? We can refer to work on the cross-cultural bouba-kiki effect [39], where people map the same shapes onto sounds. Since we perceive sounds as "soft" and "round" before conscious awareness [15], we can accept these attributes for nonvisual stimuli. Future work can explore rating studies that elicit perceptions with scales or qualitative input using the list of visual kawaii attributes to start.

### 5.2.2 Imagining Novel Qualities Specific to Kawaii Vocalics.

We also need to explore new attributes that are voice-specific. Vision and audition are distinct modalities. Kawaii vocalics may, therefore, feature unique attributes that may not be perceived in visual stimuli. For example, supplementary nonverbal kawaii sounds, like beeps and chortles, and sounds produced or recorded using bilabial mechanisms, for which Kumagai provides a starter list [20]. Exploratory work, such as qualitative perception studies, may elucidate this and seed future measures and instruments for kawaii vocalics, a crucial issue in voice UX work generally [46].

### 5.2.3  Mapping out Kawaii Voice Personas.

Kawaii is a multifaceted concept, with a variety of "flavours" available, including "kimo" or scary kawaii, "shy" and "confident" kawaii [49], Harajuku "street fashion" kawaii [36], and more. Future work can uncover the "flavours" of kawaii vocalics, especially ones that map onto certain roles or contexts of use for VAs, i.e., personas for kawaii vocalics.

### 5.2.4  Investigating Multimodality in Kawaii Voice and Speech.

Speech is the "what" of voice: the content expressed through voice as the aural medium. The relationship between voice and speech may therefore be important. If what we say is not deemed "kawaii," than will the voice not sound kawaii, even if it otherwise deemed as such? Future work can explore a variety of scripts and gibberish [2,3] to tease this out.

### 5.2.5  Exploring Multimodality in Kawaii Voice and Body.

Voice and body are intertwined [43]. Unlike Nittono and colleagues, who found a novel kawaii reaction to images of older adults smiling [29], we could not link kawaii to our older adult voices. This may be a limitation of the corpus, or it may signal that the body is key. The results for humanlikeness/artificiality suggest that a humanoid appearance may be important. Indeed, Hatsune Miku and other such voice-based characters may be as much about the visuals as the voice, which are experienced as a unit. Separating the voice from the body is one way to isolate what aspects invoke a sense of kawaii. This can be tested out in comparative, controlled work with, e.g., abstract Siri-like embodiments.

### 5.2.6  Probing Cross-Cultural Perceptions of Kawaii Voices: Are Kawaii Voices "Cute" Voices?

We did not have the bandwidth or resources to explore cross-cultural perceptions. However, previous work [5,6,18,20,49] indicates that kawaii vocalics may vary across cultural and language groups, which deserves study.

**5.3   LIMITATIONS**

As a first effort, we were limited by the number of participants, the lack of measures for kawaii vocalics, and the simulated environment (e.g., no embodied and interactive VA). Our methodology did not allow us to observe reactions, such as emotional expressions, also a limitation of previous work, e.g., [29,33]. The generalizability of these results beyond the Japanese context and language, as well as the TTS's used, will need to be explored in future work.

# 6   Conclusion

Kawaii is a transcultural phenomenon that is not only an expression of "cuteness" in visual appearance, but also an auditory property, an aspect of the vocal qualities of computer voice. Moreover, kawaii vocalics carry sociocultural implications related to age, gender, and their intersection. We have proposed an initial model of kawaii vocalics that includes gender and age as intersectional factors. We hope this work can act as a launching pad for further validation of this model and the exploration of other social and affective factors in kawaii vocalics.

**ACKNOWLEDGMENTS**

This work was funded by a Japan Society for the Promotion of Science (JSPS) Grants-in-Aid for Early Career Scientists (KAKENHI WAKATE) grant (no. 21K18005), as well as in part by a Tokyo Tech Young Investigator Engineering Award, the Tokyo Tech Young Science and Engineering Researchers Program (YSEP), and JASSO (Japan Student Services Organization). We also thank the Fonobono Research Institute, Li "Rikaku" Ge, and all members of the Aspire Lab for supporting this work.

Extended Abstracts of the 2023 CHI Conference on Human Factors in Computing Systems, Article No.: 63, 2023## References

[1] Sameera A. Abdul-Kader and J. C. Woods. 2015. Survey on chatbot design techniques in speech conversation systems. *International Journal of Advanced Computer Science and Applications* 6, 7, 72-80. https://doi.org/10.14569/IJACSA.2015.060712

[2] Alice Baird, Stina Hasse Jørgensen, Emilia Parada-Cabaleiro, Simone Hantke, Nicholas Cummins, and Björn Schuller. 2017. Perception of Paralinguistic Traits in Synthesized Voices. In *Proceedings of the 12th International Audio Mostly Conference on Augmented and Participatory Sound and Music Experiences* (AM '17: Audio Mostly 2017), 1-5. https://doi.org/10.1145/3123514.3123528

[3] Alice Baird, Stina Jørgensen, Emilia Parada-Cabaleiro, Nicholas Cummings, Simone Hantke, and Björn Schuller. 2018. The perception of vocal traits in synthesized voices: Age, gender, and human likeness. *Journal of the Audio Engineering Society* 66, 277-285. https://doi.org/10.17743/jaes.2018.0023

[4] Dave Berque, Hiroko Chiba, Ayako Hashizume, Masaaki Kurosu, and Samuel Showalter. 2019. Cuteness in Japanese design: Investigating perceptions of kawaii among American college students. In *Advances in Affective and Pleasurable Design* (International Conference on Applied Human Factors and Ergonomics (AHFE 2018)), 392-402. https://doi.org/10.1007/978-3-319-94944-4_43

[5] Dave Berque, Hiroko Chiba, Tipporn Laohakangvalvit, Michiko Ohkura, Peeraya Sripian, Midori Sugaya, Kevin Bautista, Jordyn Blakey, Feng Chen, Wenkang Huang, Shun Imura, Kento Murayama, Eric Spehlmann, and Cade Wright. 2021. Cross-cultural design and evaluation of robot prototypes based on kawaii (cute) attributes. In *Cross-Cultural Design. Applications in Cultural Heritage, Tourism, Autonomous Vehicles, and Intelligent Agents* (International Conference on Human-Computer Interaction (HCII 2021)), 319-334. https://doi.org/10.1007/978-3-030-77080-8_26

[6] Dave Berque, Hiroko Chiba, Michiko Ohkura, Peeraya Sripian, and Midori Sugaya. 2020. Fostering cross-cultural research by cross-cultural student teams: A case study related to kawaii (cute) robot design. In *Cross-Cultural Design. User Experience of Products, Services, and Intelligent Environments* (International Conference on Human-Computer Interaction (HCII 2020)), 553-563. https://doi.org/10.1007/978-3-030-49788-0_42

[7] Joy Buolamwini and Timnit Gebru. 2018. Gender shades: Intersectional accuracy disparities in commercial gender classification. In *Proceedings of the 1st Conference on Fairness, Accountability and Transparency in Machine Learning Research* (PMLR 2018), 77-91. Retrieved from http://proceedings.mlr.press/v81/buolamwini18a.html?mod=article_inline

[8] Yiqun T. Chen, Angela D. R. Smith, Katharina Reinecke, and Alexandra To. 2022. Collecting and reporting race and ethnicity data in HCI. In *CHI Conference on Human Factors in Computing Systems Extended Abstracts* (CHI '22), 1-8. https://doi.org/10.1145/3491101.3519685

[9] Adrian David Cheok. 2010. Kawaii/cute interactive media. In *Art and Technology of Entertainment Computing and Communication: Advances in Interactive New Media for Entertainment Computing*, Adrian David Cheok (ed.). Springer, London, 223-254. https://doi.org/10.1007/978-1-84996-137-0_9

[10] Leigh Clark, Philip Doyle, Diego Garaialde, Emer Gilmartin, Stephan Schlögl, Jens Edlund, Matthew Aylett, João Cabral, Cosmin Munteanu, Justin Edwards, and Benjamin R Cowan. 2019. The state of speech in HCI: Trends, themes and challenges. *Interacting with Computers* 31, 4, 349-371. https://doi.org/10.1093/iwc/iwz016

[11] Kimberlé Crenshaw. 2017. On Intersectionality: Essential Writings. *Faculty Books*. Retrieved from https://scholarship.law.columbia.edu/books/255

[12] Joe Crumpton and Cindy L. Bethel. 2016. A survey of using vocal prosody to convey emotion in robot speech. *International Journal of Social Robotics* 8, 2, 271-285. https://doi.org/10.1007/s12369-015-0329-4

[13] Leanne Hinton, Johanna Nichols, and John J. Ohala. 2006. *Sound Symbolism*. Cambridge University Press.

[14] Hsiu-Fang Hsieh and Sarah E. Shannon. 2005. Three approaches to qualitative content analysis. *Qualitative Health Research* 15, 9, 1277-1288. https://doi.org/10.1177/1049732305276687

[15] Shao-Min Hung, Suzy J. Styles, and Po-Jang Hsieh. 2017. Can a word sound like a shape before you have seen it? Sound-shape mapping prior to conscious awareness. *Psychological Science* 28, 3, 263-275. https://doi.org/10.1177/0956797616677313

[16] Yomota Inuhiko. 2006. *"Kawaii" Ron [The Theory of Kawaii]*. Chikuma Shobō, Tokyo, Japan. Retrieved December 16, 2022 from https://www.worthsharing.jpf.go.jp/en/vol_1/the-theory-of-kawaii/

[17] Lorenz Cuno Klopfenstein, Saverio Delpriori, Silvia Malatini, and Alessandro Bogliolo. 2017. The rise of bots: A survey of conversational interfaces, patterns, and paradigms. In *Proceedings of the 2017 Conference on Designing Interactive Systems* (DIS '17), 555-565. https://doi.org/10.1145/3064663.3064672

[18] Kyoko Koma. 2013. Kawaii as represented in scientific research: The possibilities of kawaii cultural studies. *Hemispheres. Studies on Cultures and Societies*, 28, 103-117.

[19] Morten L. Kringelbach, Eloise A. Stark, Catherine Alexander, Marc H. Bornstein, and Alan Stein. 2016. On cuteness: Unlocking the parental brain and beyond. *Trends in Cognitive Sciences* 20, 7, 545-558. https://doi.org/10.1016/j.tics.2016.05.003

[20] Gakuji Kumagai. 2020. The pluripotentiality of bilabial consonants: The images of softness and cuteness in Japanese and English. *Open Linguistics* 6, 1, 693-707. https://doi.org/10.1515/opli-2020-0040

[21] Neha Kumar and Naveena Karusala. 2019. Intersectional computing. *Interactions* 26, 2, 50-54. https://doi.org/10.1145/3305360

[22] Shiri Lieber-Milo. 2021. Cute at an older age: A case study of Otona-Kawaii. *Mutual Images Journal*, 10, 93-108. https://doi.org/10.32926/2021.10.lie.otona

[23] Konrad Lorenz. 1943. Die angeborenen Formen möglicher Erfahrung. *Zeitschrift für Tierpsychologie* 5, 2, 235-409. https://doi.org/10.1111/j.1439-0310.1943.tb00655.x

[24] Xingyang Lv, Yue Liu, Jingjing Luo, Yuqing Liu, and Chunxiao Li. 2021. Does a cute artificial intelligence assistant soften the blow? The impact of cuteness on customer tolerance of assistant service failure. *Annals of Tourism Research* 87, 103114. https://doi.org/10.1016/j.annals.2020.103114

[25] Aaron Marcus, Ayako Hashizume, Masaaki Kurosu, and Xiaojuan Ma. 2017. *Cuteness Engineering: Designing Adorable Products and Services*. Springer International Publishing : Imprint: Springer, Cham. https://doi.org/10.1007/978-3-319-61961-3

[26] Nicola Marsden and Monika Pröbster. 2019. Personas and identity: Looking at multiple identities to inform the construction of personas. In *Proceedings of the 2019 CHI Conference on Human Factors in Computing Systems* (CHI '19), 1-14. https://doi.org/10.1145/3290605.3300565

[27] Ryo Minakawa and Tetsuji Takada. 2017. Exploring alternative security warning dialog for attracting user attention: Evaluation of "kawaii" effect and its additional stimulus combination. In *Proceedings of the 19th International Conference on Information Integration and Web-based Applications & Services* (iiWAS '17), 582-586. https://doi.org/10.1145/3151759.3151846

[28] Hiroshi Nittono. 2010. A behavioral science framework for understanding kawaii. In *Proceedings of The Third International Workshop on Kansei*, 80-83.

[29] Hiroshi Nittono. 2016. The two-layer model of 'kawaii': A behavioural science framework for understanding kawaii and cuteness. *East Asian Journal of Popular Culture* 2, 1, 79-95. https://doi.org/10.1386/eapc.2.1.79_1

[30] Hiroshi Nittono. 2022. The psychology of "kawaii" and its implications for human-robot interaction. In *2022 17th ACM/IEEE International Conference on Human-Robot Interaction (HRI)* (HRI '22), 3-3. https://doi.org/10.1109/HRI53351.2022.9889591

[31] Hiroshi Nittono, Michiko Fukushima, Akihiro Yano, and Hiroki Moriya. 2012. The power of kawaii: Viewing cute images promotes a careful behavior and narrows attentional focus. *PLOS ONE* 7, 9, e46362. https://doi.org/10.1371/journal.pone.0046362

[32] Hiroshi Nittono and Namiha Ihara. 2017. Psychophysiological responses to kawaii pictures with or without baby schema. *SAGE Open* 7, 2, 2158244017709321. https://doi.org/10.1177/2158244017709321

[33] Hiroshi Nittono, Shiri Lieber-Milo, and Joshua P. Dale. 2021. Cross-cultural comparisons of the cute and related concepts in Japan, the United States, and Israel. *SAGE Open* 11, 1, 2158244020988730. https://doi.org/10.1177/2158244020988730

[34] Michiko Ohkura. 2019. *Kawaii Engineering: Measurements, Evaluations, and Applications of Attractiveness*. Springer, Singapore.

[35] Yuka Okada, Mitsuhiko Kimoto, Takamasa Iio, Katsunori Shimohara, Hiroshi Nittono, and Masahiro Shiomi. 2020. Can a robot's touches express the feeling of kawaii toward an object? In *2020 IEEE/RSJ International Conference on Intelligent Robots and Systems (IROS)* (IROS 2020), 11276-11283. https://doi.org/10.1109/IROS45743.2020.9340874

[36] Manami Okazaki and Geoff Johnson. 2013. *Kawaii!: Japan's Culture of Cute*. Prestel, Munich.

[37] Donnalyn Pompper. 2014. Social identities are intersectional. In *Practical and Theoretical Implications of Successfully Doing Difference in Organizations*. Emerald Group Publishing Limited, 45-61. https://doi.org/10.1108/S2051-2333(2014)0000001002

[38] Fernando Poyatos. 1993. *Paralanguage: A Linguistic and Interdisciplinary Approach to Interactive Speech and Sounds*. John Benjamins Publishing Company, Amsterdam/Philadelphia.

[39] V.S. Ramachandran and E.M. Hubbard. 2001. Synaesthesia -- A window into perception, thought and language. *Journal of Consciousness Studies* 8, 12, 3-34.

[40] Ari Schlesinger, W. Keith Edwards, and Rebecca E. Grinter. 2017. Intersectional HCI: Engaging identity through gender, race, and class. In *Proceedings of the 2017 CHI Conference on Human Factors in Computing Systems* (CHI '17), 5412-5427. https://doi.org/10.1145/3025453.3025766

[41] Howard Schuman and Stanley Presser. 1996. *Questions and Answers in Attitude Surveys: Experiments on Question Form, Wording, and Context*. SAGE, Thousand Oaks, CA, USA.

[42] Katie Seaborn and Alexa Frank. 2022. What pronouns for Pepper? A critical review of gender/ing in research. In *Proceedings of the 2022 CHI Conference on Human Factors in Computing Systems* (CHI '22), 1-15. https://doi.org/10.1145/3491102.3501996

[43] Katie Seaborn, Norihisa P. Miyake, Peter Pennefather, and Mihoko Otake-Matsuura. 2022. Voice in human–agent interaction: A survey. *ACM Computing Surveys* 54, 4, 1-43. https://doi.org/10.1145/3386867

[44] Katie Seaborn and Peter Pennefather. 2022. Neither "hear" nor "their": Interrogating gender neutrality in robots. In *Proceedings of the 2022 ACM/IEEE International Conference on Human-Robot Interaction* (HRI '22), 1030-1034. https://doi.org/10.5555/3523760.3523929

[45] Katie Seaborn, Peter Pennefather, and Haruki Kotani. 2022. Exploring gender-expansive categorization options for robots. In *Extended Abstracts of the 2022 CHI Conference on Human Factors in Computing Systems* (CHI EA '22), 1-6. https://doi.org/10.1145/3491101.3519646

[46] Katie Seaborn and Jacqueline Urakami. 2021. Measuring voice UX quantitatively: A rapid review. In

*Extended Abstracts of the 2021 CHI Conference on Human Factors in Computing Systems* (CHI EA '21), 1-8. https://doi.org/10.1145/3411763.3451712

[47] Kanako Shiokawa. 1999. Cute but deadly: Women and violence in Japanese comics. In *Themes and Issues in Asian Cartooning: Cute, Cheap, Mad, and Sexy*, John A. Lent (ed.). Bowling Green State University Popular Press, Bowling Green, OH, USA, 93-126.

[48] Shohei Sugano, Yutaka Miyaji, and Ken Tomiyama. 2013. Study of kawaii-ness in motion – Physical properties of kawaii motion of Roomba. In *Human-Computer Interaction. Human-Centred Design Approaches, Methods, Tools, and Environments* (International Conference on Human-Computer Interaction (HCII 2017)), 620-629. https://doi.org/10.1007/978-3-642-39232-0_67

[49] Jacqueline Urakami, Nan Qie, Xinyue Kang, and Pei-Luen Patrick Rau. 2021. Cultural adaptation of "kawaii" in short mobile video applications: How the perception of "kawaii" is shaped by the cultural background of the viewer and the gender of the performer. *Computers in Human Behavior Reports* 4, 100109. https://doi.org/10.1016/j.chbr.2021.100109

[50] Selma Yilmazyildiz, Werner Verhelst, and Hichem Sahli. 2015. Gibberish speech as a tool for the study of affective expressiveness for robotic agents. *Multimedia Tools and Applications* 74, 22, 9959-9982. https://doi.org/10.1007/s11042-014-2165-1

[51] Brian J. Zhang, Knut Peterson, Christopher A. Sanchez, and Naomi T. Fitter. 2021. Exploring consequential robot sound: Should we make robots quiet and kawaii-et? In *2021 IEEE/RSJ International Conference on Intelligent Robots and Systems (IROS)* (IROS 2021), 3056-3062. https://doi.org/10.1109/IROS51168.2021.9636365